\documentclass{revtex4-2}
\usepackage{textgreek}
\usepackage{graphicx}
\usepackage{amssymb,amsmath}
\usepackage{graphicx}
\usepackage{color}
\usepackage{hyperref}
\usepackage{listings}
\usepackage{natbib}
\usepackage{upgreek}
\usepackage{setspace}

\begin{document}
\title{Fine structure of second-harmonic resonances in $\chi^{(2)}$ optical microresonators}

\author{Jan Szabados$^{1}$, Nicol\'{a}s Amiune$^{1}$, Boris Sturman$^{2}$, and Ingo Breunig$^{1,3,*}$}
\affiliation{\footnotesize{
		\mbox{$^{1}$Laboratory for Optical Systems, Department of Microsystems Engineering - IMTEK, University of Freiburg, Georges-K\"ohler-Allee 102,}\\ 79110 Freiburg, Germany\\
		\mbox{$^{2}$Institute of Automation and Electrometry, Russian Academy of Sciences, Koptyug Avenue 1, 630090 Novosibirsk, Russia}\\
		\mbox{$^{3}$Fraunhofer Institute for Physical Measurement Techniques IPM, Georges-K\"ohler-Allee 301, 79110 Freiburg, Germany}}
	\\$^*$optsys@ipm.fraunhofer.de}



\begin{abstract}
Owing to the discrete frequency spectrum of whispering gallery resonators (WGRs), the resonance and phase-matching conditions for the interacting waves in the case of second-harmonic generation (SHG) cannot generally be fulfilled simultaneously. To account for this, we develop a model describing SHG in WGRs with non-zero frequency detunings at both the pump and second-harmonic frequencies. Our model predicts strong distortions of the line shape of pump and second-harmonic resonances for similar linewidths at both frequencies; for much larger linewidths at the second-harmonic frequency, this behavior is absent. Furthermore, it describes the SHG efficiency as a function of detuning. Experimentally, one can change the WGR eigenfrequencies, and thus the relative detuning between pump and second-harmonic waves by a number of means, for example electro-optically and thermally. Using a lithium niobate WGR, we show an excellent quantitative agreement for the SHG efficiency between our experimental results and the model. Also, we show the predicted distortions of the pump and second-harmonic resonances to be absent in the lithium niobate WGR, but present in a cadmium silicon phosphide WGR, as expected from the linewidths of the resonances involved.
\end{abstract}
\maketitle
\section{Introduction}
Whispering gallery resonators (WGRs) have in recent years become a well-established platform for the investigation and usage of second-order ($\chi^{(2)}$) nonlinear-optical processes such as second-harmonic generation (SHG) and optical parametric oscillation \cite{Breunig16, Strekalov16}. While SHG was realized in a WGR in 2004\cite{Ilchenko04}, it still attracts widespread interest, particularly since it was employed as an initial step for frequency comb generation in $\chi^{(2)}$ microresonators as recently as 2020\cite{Szabados20, Hendry20}. Furthermore, it was shown that by making the pump and second-harmonic light perfectly resonant, i.e.\,by maximizing the SHG efficiency, the pump power threshold for the comb generation drops below 100~\textmu W\cite{Szabados20APLP}, which can potentially be lowered to sub-\textmu W values. An analytical description of perfectly resonant SHG can be found in \cite{Sturman11}. \\
While experimentally it is easy to make sure the pump laser frequency $\nu_\mathrm{p}$ and the WGR resonance frequency $\nu_\mathrm{mp}$ are perfectly resonant, i.e.\,zero-detuned, simultaneous achievement of perfect phase-matching for SHG is not an easy task\cite{Sturman12}. Thus, one cannot generally assume the WGR resonance frequency for the second-harmonic light and the SHG frequency to coincide. An analytical description of SHG in WGRs for non-perfect phase-matching conditions can be found in \cite{Sturman12}. For frequency comb generation, however, not only non-zero second-harmonic detunings are of interest, but also the case of non-zero pump detunings is of great relevance as it allows to access different frequency comb states\cite{Ricciardi15,Leo16,Leo16pra, Mosca18,Villois19b, Smirnov20,Lobanov20}. As the comb generation thresholds coincide with the onset of internally pumped optical parametric oscillation, it is also worth noting that non-zero pump detunings were shown to significantly influence the line shape of the pump resonance in the case of optical parametric oscillation\cite{Breunig13}, providing valuable information about the loss mechanisms of this process. Hence, in this contribution, we expand the previously published model for SHG for non-perfect phase-matching, i.e.\, for non-zero second-harmonic detunings\cite{Sturman12} to also include non-zero pump detunings. In the case of zero pump detuning, we return to the results found in \cite{Sturman12}. We validate the predictions of the model using a lithium niobate (LN) WGR: making use of the linear electro-optic effect in this material, one can influence the detunings in a well-controlled manner\cite{Fuerst10shg,Szabados20APLP} and measure the corresponding SHG efficiency. As in the case of lithium niobate our model does not predict significant distortions of the pump and second-harmonic resonances, we furthermore use a cadmium silicon phosphide (CSP) WGR to observe such distortions, validating another aspect of our model.
\section{Theoretical considerations}
\begin{figure}[h]
	\begin{center}
		\includegraphics{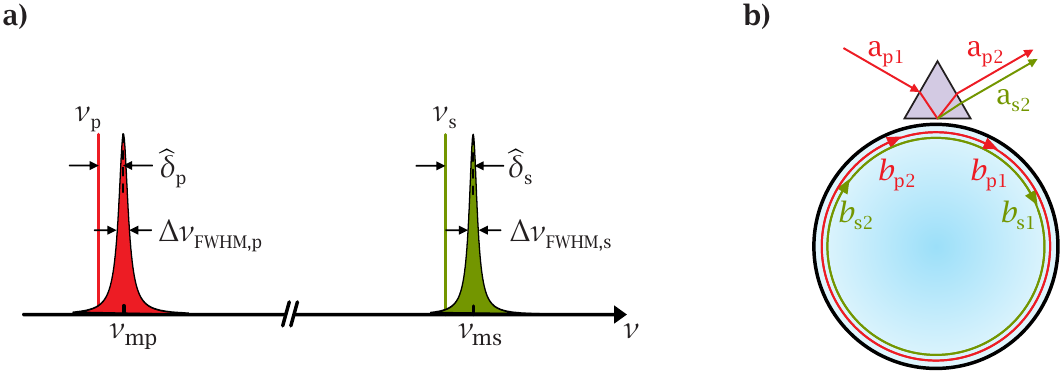}
		\caption{a) The pump (red) and second-harmonic (green) resonances in a WGR can be described by their eigenfrequencies $\nu_\mathrm{mp}$ and $\nu_\mathrm{ms}$ and their linewidths $\Delta\nu_\mathrm{FWHM,p}$ and $\Delta\nu_\mathrm{FWHM,s}$, respectively. In second-harmonic generation, pump frequency $\nu_\mathrm{p}$ is doubled to $\nu_\mathrm{s}=2\nu_\mathrm{p}$. These frequencies do not necessarily equal the eigenfrequencies $\nu_\mathrm{mp}$ and $\nu_\mathrm{ms}$. The difference between the frequencies is described by the (normalized) detunings $\hat{\delta}_\mathrm{p}$ and $\hat{\delta}_\mathrm{s}$, respectively. b) Pump light is coupled into a WGR and generates second-harmonic light inside. This situation can be described by using the complex quantities $a_{1}$ and $a_{2}$ ($b_{1}$ and $b_{2}$), which can be considered the input and output (output and input) external (internal) amplitudes. The index 'p' (red) stands for the pump, 's' (green) for the second-harmonic fields. As the second-harmonic light is generated inside the WGR, $a_\mathrm{s1}=0$.}
		\label{fig_intro}
	\end{center}	
\end{figure}

\begin{figure}[h]
	\begin{center}
		\includegraphics{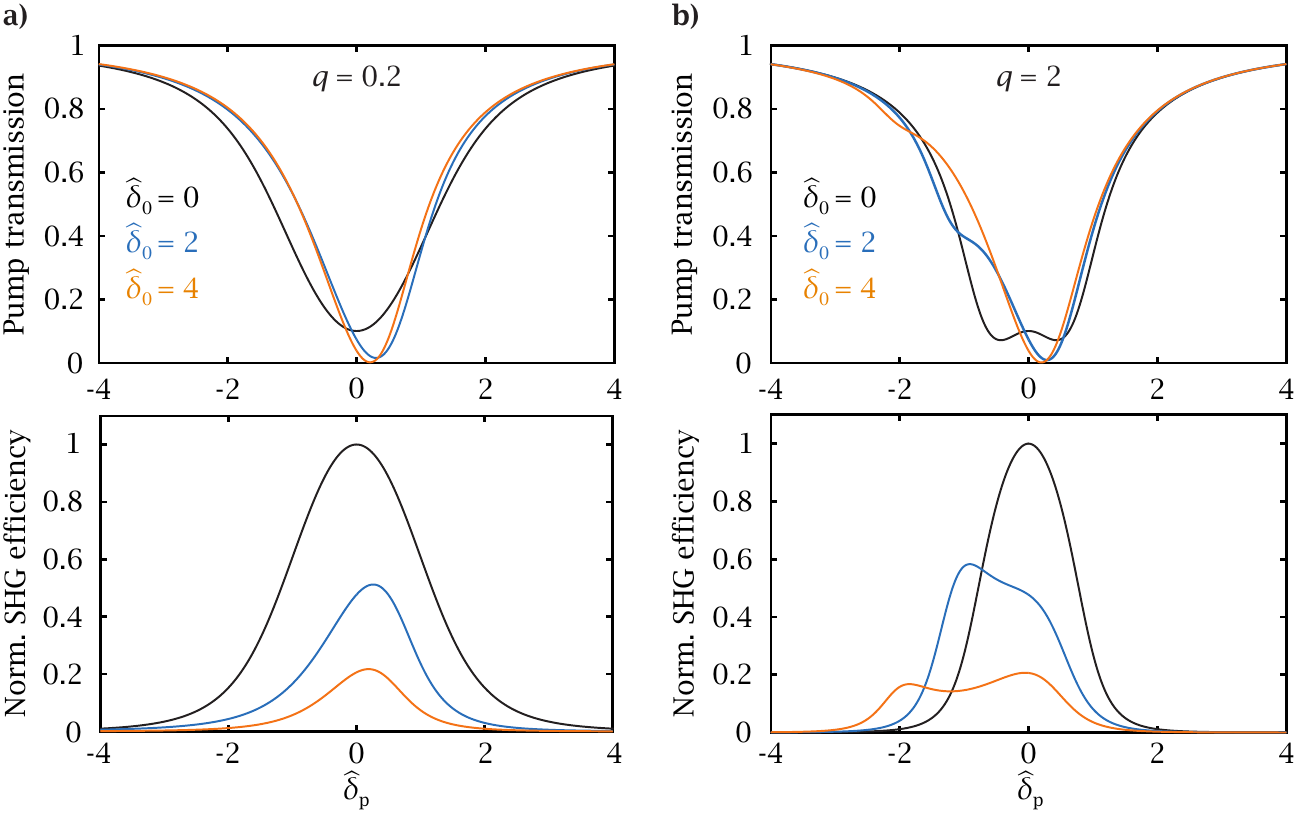}
		\caption{The pump transmission (top row) as well as the second-harmonic generation (SHG) efficiency (bottom row) assuming $r_\mathrm{s}=r_\mathrm{p}=x=1$. In a), i.e.\,for $q=0.2$, the SHG efficiency decreases with increasing $\hat{\delta}_{0}$, while the position of the second-harmonic peak is always close to $\hat{\delta}_\mathrm{p}=0$. This situation is somewhat different for b), i.e.\,for $q=2$. While the SHG efficiency similarly decreases for growing $\hat{\delta}_{0}$, one can not only observe a split of the minimum of the pump transmission for $\hat{\delta}_{0}=0$ (black curve), but also this split to be shifted away from $\hat{\delta}_\mathrm{p}=0$ for $\hat{\delta}_{0}\neq0$ (orange and blue curves). }
		\label{fig_sim}
	\end{center}	
\end{figure}

The goal of our modeling is to find an analytical relation for the SHG efficiency $\eta_\mathrm{s}$ for non-zero pump and second-harmonic frequency detunings assuming the system to be in steady state. It is based on Yariv's generic approach to the description of linear WGR phenomena\cite{Yariv2000, Yariv2002} and can be considered an extension of a previously introduced model, where non-zero frequency detunings for the second-harmonic light were considered\cite{Sturman12}. There, the pump frequency $\nu_\mathrm{p}$ and the WGR eigenfrequency $\nu_\mathrm{mp}$ are considered to be coincident (Fig.\,\ref{fig_intro}a)).
Although, as previously mentioned, experimentally this can be the case, often it is not, thus requiring a generalization of the previously mentioned model with regards to a non-zero pump detuning $\delta_\mathrm{p}\neq 0$. Analogously to \cite{Sturman12}, the detuning parameters are defined as $\delta_\mathrm{p,s}=2\pi n_\mathrm{p,s}(\nu_\mathrm{p,s}-\nu_\mathrm{mp,ms})/c_{0}$ for the pump (p) and second-harmonic (s) light, respectively, where $n_\mathrm{p,s}$ stands for the respective refractive indices and $c_{0}$ is the vacuum speed of light. Since for second-harmonic generation $\nu_\mathrm{s}=2\nu_\mathrm{p}$, we obtain the link 
\begin{equation}
\frac{\delta_\mathrm{s}}{n_\mathrm{s}}-\frac{2\delta_\mathrm{p}}{n_\mathrm{p}}=\frac{2\pi}{c_{0}}\left(2\nu_\mathrm{mp}-\nu_\mathrm{ms}\right)\equiv\delta_{0}.
\label{delta0_eq1}
\end{equation}
Thus, only one of the two non-zero detunings $\delta_\mathrm{s}$ and $\delta_\mathrm{p}$ is independent. The case considered in \cite{Sturman12} is $\delta_\mathrm{p}=0$, $\delta_\mathrm{s}=n_\mathrm{s}\delta_{0}$.\\
To obtain a formula for the SHG efficiency $\eta_\mathrm{s}$, firstly we need the coupling relations 
\begin{equation}
a_{2}=ta_{1}+\kappa b_{2},~~~~~~~~~~~~~~~~~~ b_{1}=-\kappa a_{1}+tb_{2},
\label{couplingmatrix}
\end{equation}
where $\kappa$ and $t$ are experimentally controllable real variable quantities such that $\kappa^{2}+t^{2}=1$ and $\kappa^{2}\ll 1$ and the complex quantities $b_{1}$ and $b_{2}$ ($a_{1}$ and $a_{2}$) can be considered the output and input (input and output) internal (external) amplitudes. The amplitudes are all normalized such that their absolute squared values give the modal light powers (Fig.\,\ref{fig_intro}b)). In the following, the parameters of Eq.\,(\ref{couplingmatrix}) are additionally marked by the subscripts p and s relevant to the pump and second-harmonic light, respectively. This way, the SHG efficiency can be expressed as $\eta_\mathrm{s}=\left|a_\mathrm{s2}\right|^{2}/|a_\mathrm{p1}|^{2}$, while another observable quantity, the pump transmission, is $T_\mathrm{p}=|a_\mathrm{p2}|^{2}/|a_\mathrm{p1}|^{2}$. Furthermore, obviously for the external second-harmonic input amplitude, we have $a_\mathrm{s1}=0$.\\
In the next step, we need the internal circulation relations, which account self-consistently for phase changes, extinction, and nonlinear changes of the amplitudes owing to $\chi^{(2)}$ effects. They can be expressed as
\begin{equation}
b_\mathrm{s2}=b_\mathrm{s1}\left(1-\alpha_\mathrm{s}L/2+i\delta_\mathrm{s}L\right)-i\xi Lb_\mathrm{p1}^{2},
\label{eq3_bs2}
\end{equation}
\begin{equation}
b_\mathrm{p2}=b_\mathrm{p1}\left(1-\alpha_\mathrm{p}L/2+i\delta_\mathrm{p}L\right)-i\xi Lb_\mathrm{s1}b_\mathrm{p1}^{*},
\label{eq4_bp2}
\end{equation}
where $\alpha_\mathrm{p,s}$ is the absorption coefficient of the material the WGR is made of at the respective frequencies, $L$ is the geometric circumference of said resonator (Fig.\,\ref{fig_intro}b)), and $\xi$ is the effective coupling constant as introduced in \cite{Sturman11}. While Eq.\,(\ref{eq3_bs2}) coincides with Eq.\,(8) of \cite{Sturman12}, Eq.\,(\ref{eq4_bp2}) can be considered an extension of Eq.\,(9) of \cite{Sturman12} allowing to take into account non-zero pump detunings $\delta_\mathrm{p}\neq0$.  \\
Combining Eqs.\,(\ref{eq3_bs2}) and (\ref{eq4_bp2}) with the first of Eqs.\,(\ref{couplingmatrix}) and taking into account $\kappa_\mathrm{p,s}^{2}+\alpha_\mathrm{p,s}L=2\pi/\hat{f}_\mathrm{p,s}$, where $\hat{f}_\mathrm{p,s}$ are the loaded modal finesses, we obtain after some algebraic calculations
\begin{equation}
b_\mathrm{p1}\left[1-i\hat{\delta}_\mathrm{p}+\frac{\hat{f}_\mathrm{s}\hat{f}_\mathrm{p}\xi^{2}L^{2}|b_\mathrm{p1}|^{2}}{\pi^{2}(1-i\hat{\delta}_\mathrm{s})}\right]=-\frac{\kappa_\mathrm{p}\hat{f}_\mathrm{p}a_\mathrm{p1}}{\pi}.
\label{eq7_bp1}
\end{equation}
Here, $\hat{\delta}_\mathrm{p,s}=2\hat{f}_\mathrm{p,s}R\delta_\mathrm{p,s}$ are the normalized detunings. These normalized detunings can also be expressed using frequencies as $\hat{\delta}_\mathrm{p,s}=2(\nu_\mathrm{p,s}-\nu_\mathrm{mp,ms})/\Delta\nu_\mathrm{FWHM,p,s}$ (Fig.\,\ref{fig_intro}a)), as derived in the Appendix. The above relation can be simplified by introducing the new variables 
\begin{equation}
x=\frac{\kappa_\mathrm{p}^{2}\hat{f}_\mathrm{p}^{3}\hat{f}_\mathrm{s}\xi^{2}L^{2}|a_\mathrm{p1}|^{2}}{\pi^{4}},~~~~~~~~~~~~~~~~~~ y=\frac{\pi b_\mathrm{p1}}{a_\mathrm{p1}\kappa_\mathrm{p}\hat{f}_\mathrm{p}}.
\label{eq8_xy}
\end{equation}
The real variable $x$ is the normalized external pump power, while the complex variable $y$ stands for the normalized internal pump amplitude. With these variables, Eq.\,(\ref{eq7_bp1}) changes to
\begin{equation}
y\left(1-i\hat{\delta}_\mathrm{p}+\frac{x|y|^{2}}{1-i\hat{\delta}_\mathrm{s}}\right)=-1.
\label{eq9_y}
\end{equation}
We can now express the experimentally observable SHG efficiency $\eta_\mathrm{s}$ as a function of these newly introduced quantities:
\begin{equation}
\eta_\mathrm{s}=\frac{|a_\mathrm{s2}|^{2}}{|a_\mathrm{p1}|^{2}}=\frac{1}{(1+r_\mathrm{p})(1+r_\mathrm{s})}\frac{4x|y|^{4}}{1+\hat{\delta}_\mathrm{s}^{2}},
\label{eq10_etas}
\end{equation}
where $r_\mathrm{p,s}=\kappa_\mathrm{p,s}^{2}/\alpha_\mathrm{p,s}L$ are the ratios of coupling and internal losses for the pump and the second-harmonic light, respectively. It should be noted that for zero-detunings, $\eta_\mathrm{s}$ has a maximum at $x=4$\cite{Sturman12}. For the other experimentally observable quantity, the pump transmission, we obtain
\begin{equation}
T_\mathrm{p}=\frac{|a_\mathrm{p2}|^{2}}{|a_\mathrm{p1}|^{2}}=\left|1+2y\frac{r_\mathrm{p}}{1+r_\mathrm{p}}\right|^{2}.
\label{eq11_Tp}
\end{equation}
To be able to analyze the SHG efficiency and the pump transmission, we now need to solve Eq.\,(\ref{eq9_y}). Taking its squared modulus, we get
\begin{equation}
x=\frac{1}{|y^{2}|}\left[\hat{\delta}_\mathrm{s}\hat{\delta}_\mathrm{p}-1\pm\sqrt{\frac{1+\hat{\delta}_\mathrm{s}^{2}}{|y|^{2}}-(\hat{\delta}_\mathrm{s}+\hat{\delta}_\mathrm{p})^{2}}\right].
\label{eq12_x}
\end{equation}
We see now that the allowed values of $|y|^{2}$ are restricted by the inequality $|y|^{2}\leq (1+\hat{\delta}_\mathrm{s}^{2})/(\hat{\delta}_\mathrm{s}+\hat{\delta}_\mathrm{p})^{2}$. Recalling that $\hat{\delta}_\mathrm{s}$ and $\hat{\delta}_\mathrm{p}$ are not independent (Eq.\,(\ref{delta0_eq1})), it is useful to link them via the relation 
\begin{equation}
\hat{\delta}_\mathrm{s}-q\hat{\delta}_\mathrm{p}=\hat{\delta}_\mathrm{0},
\label{eq13_qdelta0}
\end{equation}
where $q=2\hat{f}_\mathrm{s}n_\mathrm{s}/\hat{f}_\mathrm{p}n_\mathrm{p}$ and $\hat{\delta}_{0}$ can be considered experimentally controllable external parameters. These parameters can be represented in the form (see Appendix)
\begin{equation}
 q=2\frac{\Delta\nu_\mathrm{FWHM,p}}{\Delta\nu_\mathrm{FWHM,s}},
\label{eq_q}
\end{equation}
\begin{equation}
\hat{\delta}_{0}=\frac{2\nu_\mathrm{mp}-\nu_\mathrm{ms}}{\Delta\nu_\mathrm{FWHM,s}/2},
\label{eq_expdelta0}
\end{equation}
where $\Delta\nu_\mathrm{FWHM,p,s}=\Delta\nu_\mathrm{FWHM,0p,0s}(1+r_\mathrm{p,s})$ are the linewidths of the cavity resonances at the pump and second-harmonic frequencies, respectively, and $\Delta\nu_\mathrm{FWHM,0p,0s}$ are the intrinsic linewidths.
Using Eqs.\,(\ref{eq9_y})-(\ref{eq13_qdelta0}), we can now calculate the SHG efficiency $\eta_\mathrm{s}$ and the pump transmission $T_\mathrm{p}$ for different values of $q$ and $\hat{\delta}_{0}$ against the pump detuning $\hat{\delta}_\mathrm{p}$. This is illustrated in Fig.\,\ref{fig_sim}. 
For the examples shown in the figure, for the sake of simplicity we choose $r_\mathrm{p}=r_\mathrm{s}=x=1$. It should be noted, however, that these particular values do not change the qualitative behavior that is under investigation. If the finesses $\hat{f}_\mathrm{p}$ and $\hat{f}_\mathrm{s}$ are the same, it is $q\simeq2$ (Fig.\,\ref{fig_sim}b)). In many experiments considering SHG, however, $\hat{f}_\mathrm{s}$ is considerably smaller than $\hat{f}_\mathrm{p}$, thus leading to substantially smaller values for $q$\cite{Fuerst10shg, Fuerst15,Szabados20, Szabados20APLP}. To account for this case as well, we calculated $\eta_\mathrm{s}$ and $T_\mathrm{p}$ for $q=0.2$ (Fig.\,\ref{fig_sim}a)). \\
Perhaps unsurprisingly, $\eta_\mathrm{s}$ decreases with growing $|\hat{\delta}_{0}|$ for both values of $q$. Interestingly, however, one can see a qualitative difference between the two cases: while for $q=0.2$, the position of the maximum of $\eta_\mathrm{s}$ is always at $\hat{\delta}_\mathrm{p}\simeq 0$ regardless of $|\hat{\delta}_{0}|$, for $q=2$ it significantly shifts the position of the maximum. Changing the sign of $\hat{\delta}_{0}$ results in a reflection of the resulting curve about the vertical axis. Apart from the shift of the position of the maximum, one can also observe that the shape of the SHG efficiency curves changes significantly for $q=2$. This effect is also visible in the pump transmission: while for $q=0.2$, the curves clearly show one dip only, which is at the same position as the corresponding maximum of $\eta_\mathrm{s}$, for $q=2$ one can observe a structural change in the pump transmission where the maximum of $\eta_\mathrm{s}$ lies; in this case, one can see an offset between the global minimum of $T_\mathrm{p}$ and the maximum of $\eta_\mathrm{s}$, a feature that cannot be observed for $q=0.2$. Furthermore, for $\hat{\delta}_{0}=0$, one can observe a split of the minimum of $T_\mathrm{p}$ for $q=2$, while this feature is not present for $q=0.2$. Our analysis shows that this split is present for $q\gtrsim 0.6$. For $\hat{\delta}_{0}\neq 0$, one can consider the split to be shifted away from $\hat{\delta}_\mathrm{p}=0$. 
 \subsection{Application of the model to lithium niobate}
 To validate the model experimentally, we need to find a way to control $\hat{\delta}_{0}$ (Eq.\,(\ref{eq_expdelta0})). This can be done by changing $2\nu_\mathrm{mp}-\nu_\mathrm{ms}$, i.e.\,by changing the WGR eigenfrequencies . In our material of choice, lithium niobate (LN), a convenient way of changing the eigenfrequencies is provided by the linear electro-optic effect, the so-called Pockels effect. Here, the eigenfrequencies are changed linearly by applying an external electric field $E$\cite{Fuerst10shg}. Generating second-harmonic light via birefringent phase-matching, the pump and the second-harmonic light have different polarizations\cite{Fuerst10shg}. This as well as their different frequencies results in them experiencing different electro-optic coefficients $\tilde{r}_\mathrm{p,s}$, respectively, leading to eigenfrequency changes
 \begin{equation}
 \Delta\nu_\mathrm{p,s}(E)=\nu_\mathrm{mp,ms}(E)-\nu_\mathrm{mp,ms}(0)=\frac{1}{2}\nu_\mathrm{mp,ms}n_\mathrm{p,s}^{2}\tilde{r}_\mathrm{p,s}E.
 \label{eq_eigfreqchange}
 \end{equation}
 Assuming without loss of generality $\hat{\delta}_{0}(0)=0$ and combining Eqs.\,(\ref{eq_expdelta0}) and (\ref{eq_eigfreqchange}), we thus obtain
 \begin{equation}
 \hat{\delta}_{0}(E)=\frac{2\nu_\mathrm{mp}(0)E\left[\tilde{r}_\mathrm{p}n_\mathrm{p}^{2}-\tilde{r}_\mathrm{s}n_\mathrm{s}^{2}\right]}{\Delta\nu_\mathrm{FWHM,s}}.
 \label{eq_delta0change}
 \end{equation}
 For realistic values of $n_\mathrm{p}=n_\mathrm{s}=2$, $\nu_\mathrm{mp}=300$~THz\cite{Breunig16}, $\tilde{r}_\mathrm{s}\approx 8$~pm/V, $\tilde{r}_\mathrm{p}\approx 3\tilde{r}_\mathrm{s}$\cite{Minet2020} and $\Delta\nu_\mathrm{FWHM,s}=10$~MHz\cite{Fuerst10shg}, we can estimate a resulting $\hat{\delta}_{0}\approx 4E$/(V/mm). Thus, by applying small bias voltages, we can easily investigate the effect of $\hat{\delta}_{0}$ on the resulting SHG efficiency $\eta_\mathrm{s}=P_\mathrm{s}/P_\mathrm{p,in}$ for a large range of values, where $P_\mathrm{s}$ is the generated second-harmonic power, while $P_\mathrm{p,in}$ stands for the power of the incoupled pump light.  
 \subsection{Application of the model to cadmium silicon phosphide}
 Analogously to the situation explained in the previous subsection for LN, we also need to find a way to experimentally control $\hat{\delta}_{0}$ in cadmium silicon phosphide (CSP). As the Pockels effect has thus far not been observed in this material, we change the eigenfrequencies by changing the temperature of the WGR. This way, one changes the refractive indices of the material according to its Sellmeier equation\cite{Wei18}. Using an analytical formula for the determination of the WGR eigenfrequencies\cite{Gorodetsky06}, we can then determine $2\nu_\mathrm{mp}-\nu_\mathrm{ms}$ and thus $\hat{\delta}_{0}$ (Eq.\,(\ref{eq_expdelta0})) for any temperature. Taking as an example the fundamental mode for both the pump and the second-harmonic, we can estimate that a shift of $\Delta T=1$~K from the perfectly resonant phase-matching temperature results in $2\nu_\mathrm{mp}-\nu_\mathrm{ms} \approx 450~\text{MHz}$. Then, with $\Delta\nu_\mathrm{FWHM,0s}\approx55$~MHz\cite{Jia18} we get  $\hat{\delta}_{0} \approx 16$, i.e.\,$\hat{\delta}_{0}\approx16\Delta T$/K. 
\section{Experimental methods}
\begin{figure}[h]
	\begin{center}
		\includegraphics{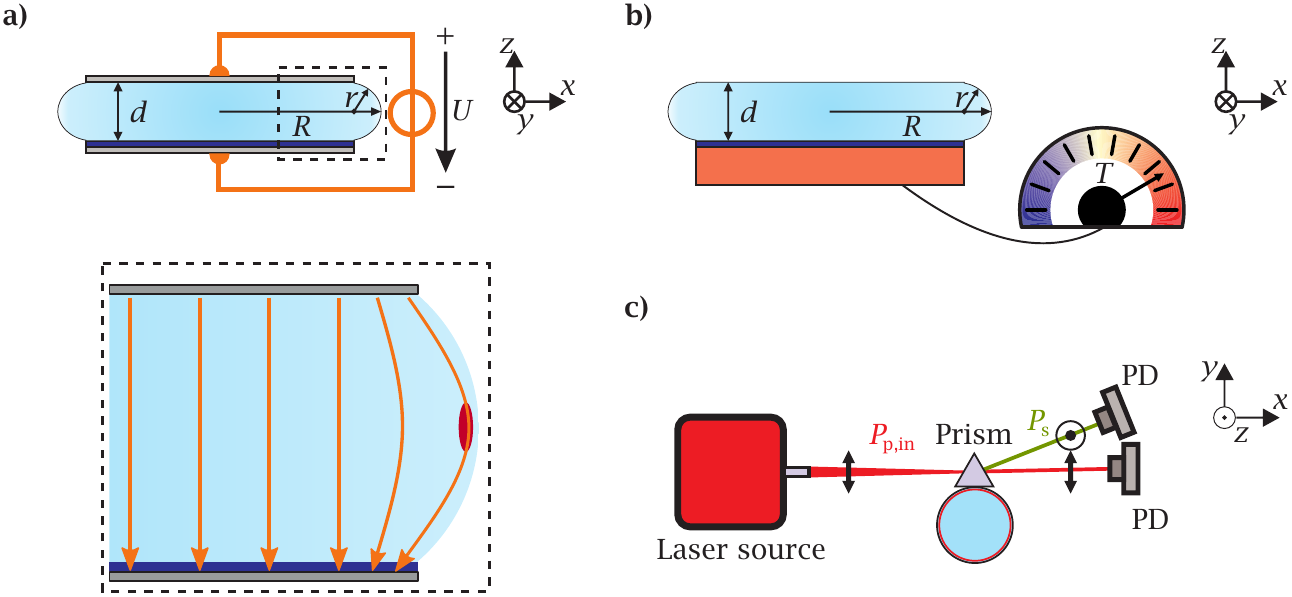}
		\caption{a) Side-view of the LN whispering gallery resonator (WGR) with thickness $d=300$~\textmu m, major radius $R=1$~mm and minor radius $r=380$~\textmu m. The resonator has a chromium electrode on top and is glued (blue layer) to a metal post, which serves as the bottom electrode. A close-up of the section indicated by the dotted lines shows a sketch of the electric field distribution (orange lines). The red area indicates the cross section and position of light traveling inside the resonator (not to scale). b) For CSP, the WGR is characterized by $d=1$~mm, $R=520$~\textmu m and $r=200$~\textmu m. It is glued to a metal post and can be heated to temperatures up to $T=100~^{\circ}$C. c) Sketch of the experimental setup. Light from a laser source is prism-coupled into a WGR. The pump (shown in red) and the generated second-harmonic (shown in green) light are spatially separated and focused onto photodetectors to measure their powers and thus the SHG efficiency.}
		\label{fig_setup}
	\end{center}	
\end{figure}

In our experiment, we generate second-harmonic light by coupling light into whispering gallery resonators (WGR) made out of 5\% MgO-doped congruent $z$-cut lithium niobate (LN) and cadmium silicon phosphide (CSP), respectively. In both cases, we start out with a piece of crystal of thickness $d$, from which we cut out a cylinder using a femtosecond laser source emitting light at 388~nm wavelength with 2~kHz repetition rate and 300~mW average output power. In the next step, the cylinder we cut out is glued to a metal post for easier handling. Subsequently, the same femtosecond laser source is used to give the resonator its desired geometry of a major radius $R$ and a minor radius $r$ as shown in Figs.\,\ref{fig_setup}a-b). To reach an optical-grade surface quality, the WGRs are eventually polished with a diamond slurry.
In the case of the LN WGR, our geometry is defined by $d=300$~\textmu m, $R=1$~mm and $r=380$~\textmu m. As previously described, we need to be able to apply bias voltages $U$ to the WGR, generating electric fields $E$ inside of it and thus changing $\hat{\delta}_{0}$ (Eq.\,(\ref{eq_delta0change})). Hence, we deposit chromium on the $+z$-side of the LN with the $-z$-side glued to the metal post (Fig.\,\ref{fig_setup}a)). This metal post acts as a bottom electrode. \\
For the CSP WGR, we change $\hat{\delta}_{0}$ by modifying the temperature. This is why the metal post is attached to a temperature controller allowing us to heat it to temperatures up to 100$~^{\circ}$C with mK-stability (Fig.\,\ref{fig_setup}b)).\\
For our experiments, we employ the same basic setup shown in Fig.\,\ref{fig_setup}c): light emitted from a laser source is focused onto a prism, which is in close proximity to the rim of the WGR; this way, we can couple light into it. First, we characterize the respective WGRs at very low pump powers to avoid thermal and nonlinear-optical effects on the linewidth: this way, we can determine the intrinsic linewidth $\Delta\nu_\mathrm{FWHM,0p}$ and the coupling efficiency. In a second step, we heat the resonator to fulfill the phase-matching condition for SHG: as we use $z$-cut LN and CSP, the pump light needs to be polarized in the $x$-$y$-plane of the crystal, thus experiencing the ordinary (o-) refractive index of the respective material. This leads to the generation of extraordinarily (e-) polarized second-harmonic light\cite{Fuerst10shg}. We separate the outcoupled o-polarized pump from the e-polarized second-harmonic light and measure both using calibrated photodetectors. This allows us to not only measure the transmission spectrum of the pump light and the second-harmonic signal, but also their respective powers and thus the SHG efficiency $\eta_\mathrm{s}$.  \\
In the case of LN, we use a frequency-tunable fiber-coupled continuous-wave laser source emitting at 1064~nm wavelength (NKT Koheras Basik, 20~kHz linewidth) passing an optical attenuator (Thorlabs~V1000A) allowing us to set the power and a polarization controller to set the polarization of the pump light. To fulfill the phase-matching conditions for SHG, we heat the WGR to $T\approx70~^{\circ}$C. Conveniently, the large birefringence of the rutile prism we employ automatically separates the pump and second-harmonic light spatially. The photodetectors for both the pump and the second-harmonic light are made of silicon. It should be noted that throughout the experiment, we stayed at pump powers about an order of magnitude below the threshold for internally pumped optical parametric oscillation as determined in \cite{Szabados20APLP}, i.e.\,$x\simeq0.4$. By applying a number of different voltages (thus changing $\hat{\delta}_{0}$) and measuring the resulting $\eta_\mathrm{s}$, we can experimentally validate the model introduced in the previous section.\\
In the case of CSP, the light sources are a fiber-coupled distributed feedback laser diode emitting light at 1.57~\textmu m wavelength and an optical parametric oscillator pumped by a Ti:sapphire laser, which provides idler light between 1.7 and 3.5~\textmu m\cite{Leidinger15}. To generate second-harmonic light, we heat the WGR to $T\approx 57.5~^{\circ}$C and couple o-polarized light with 3.14~\textmu m wavelength into it using a silicon prism. As silicon is not birefringent, the outcoupled pump and second-harmonic beams are separated with a dielectric mirror. We monitor the pump transmission and second-harmonic power with a HgCdTe detector (Vigo PVMI-4TE) and an InGaAs detector (Thorlabs PDA10DT) respectively. By measuring the second-harmonic signal for a number of different temperatures (thus changing $\hat{\delta}_{0}$), we can experimentally validate the model introduced above.
\section{Results and discussion}
\begin{figure}[h]
	\begin{center}
		\includegraphics{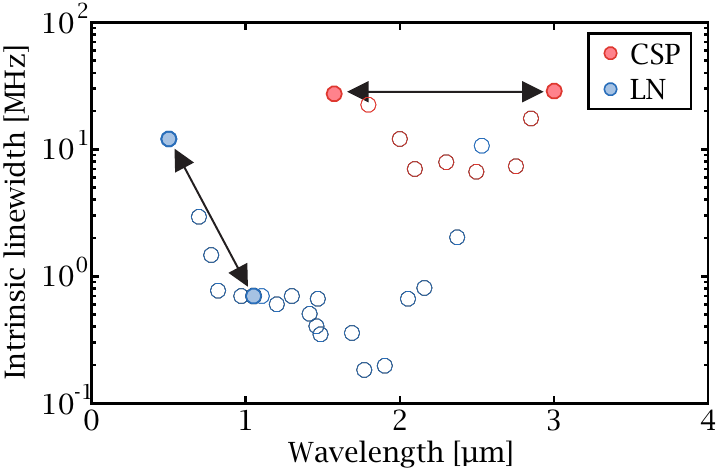}
		\caption{Intrinsic linewidths vs.\,wavelength of LN (blue) and CSP (red) WGRs. The values for the LN WGR are taken from \cite{Leidinger15}. In both cases, the linewidths are displayed for e-polarized light. The filled dots connected by arrows show the linewidths closest to the respective second-harmonic generation processes used in our experiment.}
		\label{fig_Q}
	\end{center}	
\end{figure}

\begin{figure}[h]
	\begin{center}
		\includegraphics{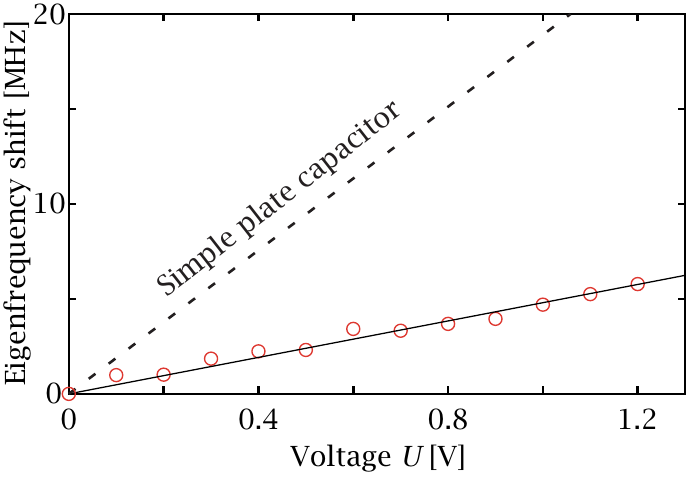}
		\caption{Change of the pump resonance eigenfrequency vs.\,applied bias voltage. The experimentally measured values are marked by the red circles. Our experimental values are described by the solid black line, indicating a reduced electric field inside the WGR compared to a simple plate capacitor.}
		\label{fig_gamma}
	\end{center}	
\end{figure}

\begin{figure}[h]
	\begin{center}
		\includegraphics{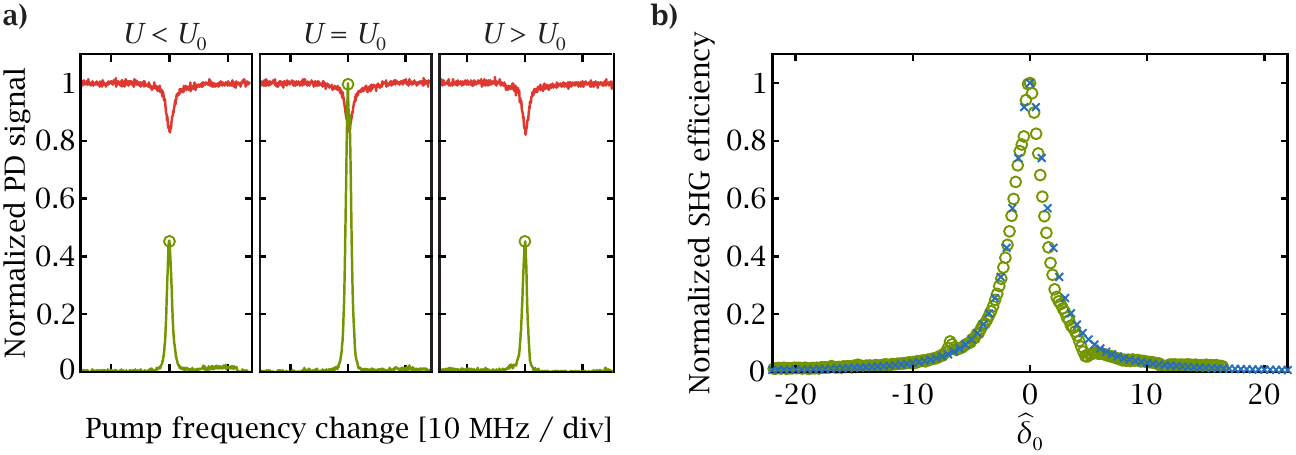}
		\caption{a) By changing the applied bias voltage $U$ to the LN WGR (and thus the detuning $\hat{\delta}_{0}$), we not only shift the pump resonances (red), but also the SHG efficiency (green curves). For $U=U_{0}$, we reach a maximum: here, $\hat{\delta}_{0}=0$. The peak values are marked by green circles. By carrying out this measurement for a number of different applied voltages (step size $\Delta U=100$~mV), we obtain the curve shown in b). Comparing this with our model for $q=0.2$ and $x\approx0.6$, we observe an excellent match between our experimental measurements (green circles) and the predictions of our model (blue crosses).}
		\label{fig_res}
	\end{center}	
\end{figure}

\begin{figure}[h]
	\begin{center}
		\includegraphics{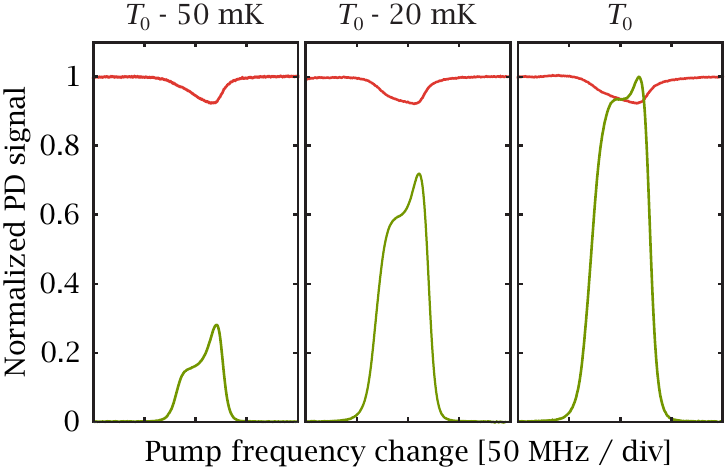}
		\caption{By changing the temperature $T$ of the CSP WGR (and thus the detuning $\hat{\delta}_{0}$), we not only shift the pump resonances (red), but also change the SHG efficiency (green curves).}
		\label{fig_res_CSP}
	\end{center}	
\end{figure}

Let us compare the two different situations for LN and CSP: when looking at their intrinsic linewidths (Fig.\,\ref{fig_Q}), one can see that while for LN $\Delta\nu_\mathrm{FWHM,0p}/\Delta\nu_\mathrm{FWHM,0s}\approx20$, the same value for CSP is $\Delta\nu_\mathrm{FWHM,0p}/\Delta\nu_\mathrm{FWHM,0s}\approx 1$. 
Owing to this, the respective values for $q$ (Eq.\,(\ref{eq_q})) are about an order of magnitude different. The linewidths are displayed for extraordinarily polarized light, respectively, but in the wavelength ranges of interest the difference between extraordinarily and ordinarily polarized light is marginal in CSP according to our measurements as well as in LN according to \cite{Leidinger15}. In LN, as $\alpha_\mathrm{s}=10\alpha_\mathrm{p}$\cite{Leidinger15} and $\kappa_\mathrm{s}<\kappa_\mathrm{p}$ for the wavelengths involved\cite{Szabados20APLP}, $r_\mathrm{s}<0.1r_\mathrm{p}$ has to be the case since $r_\mathrm{p,s}=\kappa_\mathrm{p,s}^{2}/\alpha_\mathrm{p,s}L$ as mentioned previously. As we carry out our experiment close to critical coupling, where $r_\mathrm{p}\simeq 1$\cite{Breunig16}, we can estimate $r_\mathrm{s}<0.1$. Consequently, as we know the relation $\Delta\nu_\mathrm{FWHM,p,s}=\Delta\nu_\mathrm{FWHM,0p,0s}(1+r_\mathrm{s})$, it is reasonable to assume $\Delta\nu_\mathrm{FWHM,p}=2\Delta\nu_\mathrm{FWHM,0p}$ and $\Delta\nu_\mathrm{FWHM,s}\approx\Delta\nu_\mathrm{FWHM,0s}$, thus resulting in $q\simeq0.2$ (Eq.\,(\ref{eq_q})), which is exactly what we displayed in Fig.\,\ref{fig_sim}a).\\
In the case of CSP, one can assume $r_\mathrm{p}\approx r_\mathrm{s}$\cite{Jia18} and thus $\Delta\nu_\mathrm{FWHM,p}\approx\Delta\nu_\mathrm{FWHM,s}$: this results in $q\simeq2$, which is the situation shown in Fig.\,\ref{fig_sim}b).\\
~\\
Let us now compare our experimental measurements for the respective materials to the predictions made by our model. For LN, to be able to convert from applied voltages $U$ to the relevant detuning $\hat{\delta}_{0}$, we need to determine the electric fields $E$ in the WGR (Eq.\,(\ref{eq_expdelta0})). To do this, we insert $E=\gamma U/d$ into Eq.\,(\ref{eq_eigfreqchange}). In a simple plate capacitor model, we would have $\gamma=1$: this is shown as the dashed line. In reality, however $\gamma<1$ is the case due to the WGR geometry\cite{Minet2020}; a further reduction can be expected due to the glue below the WGR. Measuring the eigenfrequency change $\Delta\nu_\mathrm{mp}$ against the applied voltage, we end up with the curve shown in Fig.\,\ref{fig_gamma}.
Inserting $\nu_\mathrm{mp}=281.749$~THz, $n_\mathrm{p}=2.2291$\cite{Umemura16}, and $\tilde{r}_\mathrm{p}=31.8$~pm/V\cite{Mendez99} into Eq.\,(\ref{eq_eigfreqchange}), eventually, we end up with $\gamma=0.25$, giving us a very good match between the experimental result and the expected linear behavior (Fig.\,\ref{fig_setup}c)). For the second-harmonic light, we know $n_\mathrm{s}=2.2263$\cite{Umemura16} and $\tilde{r}_\mathrm{s}=8.1$~pm/V\cite{Mendez99}; now we only need $\Delta\nu_\mathrm{FWHM,s}$ to obtain our values for $\hat{\delta}_{0}$ (Eq.\,(\ref{eq_expdelta0})). In our experiment, we measure $\Delta\nu_\mathrm{FWHM,0p}=1.6$~MHz. As previously discussed, it is reasonable to assume $\Delta\nu_\mathrm{FWHM,s}=20\Delta\nu_\mathrm{FWHM,0p}=32$~MHz. \\
Now that we have all the required parameters to calculate $\hat{\delta}_{0}$ (Eq.\,(\ref{eq_expdelta0})) from the applied bias voltage $U$ available, let us turn our attention to our experimental results. Changing the applied bias voltage, we don't only shift the pump resonance, but also change the SHG efficiency $\eta_\mathrm{s}$ as can be seen in Fig.\,\ref{fig_res}a).
The position of the SHG peak and the pump transmission dip coincide for all measurements, i.e.\,the maximum of the SHG efficiency is at $\hat{\delta}_\mathrm{p}\simeq0$. This behavior is as expected given that $q\simeq0.2$ (Fig.\,\ref{fig_sim}a)). We reach a maximum of $\eta_\mathrm{s}$ at a voltage $U=U_\mathrm{0}$. This is where we have $\hat{\delta}_\mathrm{p}=\hat{\delta}_\mathrm{s}=0$\cite{Szabados20APLP}, consequently leading to $\hat{\delta}_{0}=0$ according to Eq.\,(\ref{eq13_qdelta0}). Changing the applied voltage in steps of $\Delta U=100$~mV, resulting in a change of $\Delta\hat{\delta}_{0}\approx 0.18$, we can then plot the measured SHG efficiencies against $\hat{\delta}_{0}$ as visualized in Fig.\,\ref{fig_res}b). We cover a range from $\hat{\delta}_{0}=-20$ to $\hat{\delta}_{0}=15$. One can observe a decrease of the SHG efficiency against $\hat{\delta}_{0}$ that is symmetric about the vertical axis.
For $x\approx0.6$, we see that our experimental results for the SHG efficiency against $\hat{\delta}_{0}$ are described by the model very well (Fig.\,\ref{fig_res}b)) taking the maximum SHG efficiency for each $\hat{\delta}_{0}$ into account. In a previous publication, we showed that one can observe internally pumped OPO for $x=4$, as this is where the SHG efficiency peaks for the case of $\hat{\delta}_\mathrm{p}=\hat{\delta}_\mathrm{s}=0$\cite{Szabados20APLP}. As mentioned in the previous section, to only observe second-harmonic generation without any further processes, we thus made sure to stay about an order of magnitude below this threshold; hence, $x\approx0.6$ can be considered a reasonable assumption. \\
~\\
For CSP, we change $\hat{\delta}_{0}$ by changing the temperature. At a certain temperature $T=T_{0}$, we reach a maximum of the SHG efficiency as depicted in Fig.\,\ref{fig_res_CSP}: this is where we assume $\hat{\delta}_{0}\simeq0$. Changing the temperature from this value leads to a decrease in SHG efficiency as expected.
More interestingly, however, one can notice a fundamentally different behavior when it comes to the shape of the pump transmission and second-harmonic signals compared to the situation in LN (Fig.\,\ref{fig_res}a)). While the general triangular shape of the pump transmission is a thermal effect due to the large incoupled pump power\cite{Ilchenko92,Carmon04}, the double-peak structure of the second-harmonic signal strongly resembles the behavior predicted by our model for $\hat{\delta}_{0}\neq0$ as shown in Fig.\,\ref{fig_sim}b). This is in line with the predictions of our model for $q=2$, which is the value at hand for our SHG process in CSP. It should be pointed out, however, that we only observe a qualitatively similar behavior to the predictions of the model here; a more thorough investigation is needed to be able to carry out a quantitative comparison.
\section{Conclusion}
We introduce a generic model for second-harmonic generation in whispering gallery resonators taking non-zero detunings for both the pump and the second-harmonic light into account. With this model, we fully describe the SHG efficiency as well as the line shape of the pump and second-harmonic resonances as a function of experimentally accessible detunings. Using a standard $\chi^{(2)}$ material, lithium niobate, we investigate the influence of the detuning on the SHG efficiency: this way, we show the model to describe the experimentally measured second-harmonic generation efficiency against the combined detuning $\hat{\delta}_{0}$ very well for realistic values. The line shapes at both the pump and second-harmonic are not qualitatively distorted. This is different for CSP, where we observe significantly distorted line shapes for the measured pump and second-harmonic curves compared to the situation in LN. This is in line with the predictions of our model and can be explained by looking at the linewidths at the pump and second-harmonic resonances, respectively: while in LN, the second-harmonic resonance has a much larger linewidth compared to the pump resonance, in CSP they are equal. Thus, this model not only allows a way to quantitatively predict the behavior of the SHG efficiency against experimentally accessible detunings, but furthermore allows to estimate the second-harmonic linewidth, a measure for the losses of the WGR, from the measured line shapes only.  \\ While this model only describes second-harmonic generation, it can still be considered a fundamental step towards understanding $\chi^{(2)}$ frequency comb generation via SHG better, as for those combs it was shown that the detuning $\hat{\delta}_{0}$ plays a major role in accessing different comb states such as stable Turing roll patterns\cite{Szabados20}. Extending this model to contain further $\chi^{(2)}$ nonlinear-optical processes may lead to a better understanding of the generation of $\chi^{(2)}$ frequency combs, a very fast-growing field\cite{Ricciardi20}.

\section*{Funding}
Horizon 2020 Framework Programme (812818, MICROCOMB); Fraunhofer and Max Planck Cooperation Programme (COSPA).
\section*{Acknowledgments}
The authors thank Karsten Buse and Yannick Minet for helpful discussions as well as Peter G. Schunemann and Kevin T. Zawilski of BAE Systems for providing the cadmium silicon phosphide samples. 

\section*{Disclosures}
The authors declare no conflicts of interest.

\section*{Data availability} Data underlying the results presented in this paper are available from the corresponding author upon reasonable request.

\section*{Appendix}
We derive Eqs.\,(\ref{eq_q}) and (\ref{eq_expdelta0}), starting out with their description given by our model. It is important to note that the loaded finesses $\hat{f}_\mathrm{p,s}$ can be written as
\begin{equation}
\hat{f}_\mathrm{p,s}=\frac{\Delta\nu_\mathrm{FSR,p,s}}{\Delta\nu_\mathrm{FWHM,p,s}},
\label{app1}
\end{equation}
using relations given in \cite{Sturman12, Breunig16}. As from \cite{Sturman12}, we also know $\Delta\nu_\mathrm{FSR,p,s}=c_{0}/(n_\mathrm{p,s}L)$, we can re-write Eq.\,(\ref{app1}) to
\begin{equation}
\hat{f}_\mathrm{p,s}n_\mathrm{p,s}=\frac{c_{0}}{L\Delta\nu_\mathrm{FWHM,p,s}}.
\label{app2}
\end{equation} 
Knowing this, we turn our attention to $q$:
\begin{equation}
q=2\frac{\hat{f}_\mathrm{s}n_\mathrm{s}}{\hat{f}_\mathrm{p}n_\mathrm{p}}\stackrel{\mathrm{Eq.}\,(\ref{app2})}{=}2\frac{\Delta\nu_\mathrm{FWHM,p}}{\Delta\nu_\mathrm{FWHM,s}},
\end{equation}
which is exactly what's given in Eq.\,(\ref{eq_q}). \\
~\\
Let's turn to $\hat{\delta}_{0}$: starting out at Eq.\,(\ref{eq13_qdelta0}), we have
\begin{equation}
\hat{\delta}_{0}=\hat{\delta}_\mathrm{s}-q\hat{\delta}_\mathrm{p}=2\hat{f}_\mathrm{s}R\delta_\mathrm{s}-2\frac{\hat{f}_\mathrm{s}n_\mathrm{s}}{\hat{f}_\mathrm{p}n_\mathrm{p}}(2\hat{f}_\mathrm{p}R\delta_\mathrm{p})\stackrel{\mathrm{Eq.}\,(\ref{delta0_eq1})}{=}2\hat{f}_\mathrm{s}R\delta_\mathrm{s}-2\hat{f}_\mathrm{s}R\delta_\mathrm{s}\left[\frac{\delta_\mathrm{s}}{n_\mathrm{s}}-\delta_{0}\right]=2\hat{f}_\mathrm{s}Rn_\mathrm{s}\delta_{0}.
\label{appendixeq}
\end{equation}
Making use of the relations $\hat{f}_\mathrm{s}=2\pi/(\alpha_\mathrm{s}L(1+r_\mathrm{s}))$\cite{Sturman12}, $\Delta\nu_\mathrm{FWHM,s}=c_{0}\alpha_\mathrm{s}(1+r_\mathrm{s})/(2\pi n_\mathrm{s})$\cite{Breunig16}, and $L=2\pi R$, Eq.\,(\ref{appendixeq}) can be further simplified to 
\begin{equation}
\hat{\delta}_{0}=2Rn_\mathrm{s}\delta_{0}\frac{2\pi}{\alpha_\mathrm{s}L(1+r_\mathrm{s})}=\frac{2c_{0}R\delta_{0}}{L\Delta\nu_\mathrm{FWHM,s}}\stackrel{\mathrm{Eq.}\,(\ref{delta0_eq1})}{=}\frac{4\pi R(2\nu_\mathrm{mp}-\nu_\mathrm{ms})}{L\Delta\nu_\mathrm{FWHM,s}}=\frac{2\nu_\mathrm{mp}-\nu_\mathrm{ms}}{\Delta\nu_\mathrm{FWHM,s}/2},\label{eq:norm_delta0}
\end{equation}
which is exactly the same as the relation given in Eq.\,(\ref{eq_expdelta0}).\\
~\\
Finally, we will show how the normalized detunings $\hat{\delta}_\mathrm{p,s}=2\hat{f}_\mathrm{p,s}R\delta_\mathrm{p,s}$ can be expressed using frequencies only (Fig.\,\ref{fig_intro}a)). For this, we have to use the detuning parameters $\delta_\mathrm{p,s}=2\pi n_\mathrm{p,s}(\nu_\mathrm{p,s}-\nu_\mathrm{mp,ms})/c_{0}$ analogously to \cite{Sturman12}. Using this relation, we get
\begin{equation}
\hat{\delta}_\mathrm{p,s}=2\hat{f}_\mathrm{p,s}R\delta_\mathrm{p,s}=4\pi R\frac{\nu_\mathrm{p,s}-\nu_\mathrm{mp,ms}}{c_{0}}\hat{f}_\mathrm{p,s}n_\mathrm{p,s}\stackrel{\mathrm{Eq.}\,(\ref{app2})}{=}2\frac{\nu_\mathrm{p,s}-\nu_\mathrm{mp,ms}}{\Delta\nu_\mathrm{FWHM,p,s}}
\end{equation}
just as described previously.

\bibliography{paperbib_arxiv}

\providecommand{\noopsort}[1]{}\providecommand{\singleletter}[1]{#1}%
\begin{thebibliography}{30}%
\makeatletter
\providecommand \@ifxundefined [1]{%
 \@ifx{#1\undefined}
}%
\providecommand \@ifnum [1]{%
 \ifnum #1\expandafter \@firstoftwo
 \else \expandafter \@secondoftwo
 \fi
}%
\providecommand \@ifx [1]{%
 \ifx #1\expandafter \@firstoftwo
 \else \expandafter \@secondoftwo
 \fi
}%
\providecommand \natexlab [1]{#1}%
\providecommand \enquote  [1]{``#1''}%
\providecommand \bibnamefont  [1]{#1}%
\providecommand \bibfnamefont [1]{#1}%
\providecommand \citenamefont [1]{#1}%
\providecommand \href@noop [0]{\@secondoftwo}%
\providecommand \href [0]{\begingroup \@sanitize@url \@href}%
\providecommand \@href[1]{\@@startlink{#1}\@@href}%
\providecommand \@@href[1]{\endgroup#1\@@endlink}%
\providecommand \@sanitize@url [0]{\catcode `\\12\catcode `\$12\catcode
  `\&12\catcode `\#12\catcode `\^12\catcode `\_12\catcode `\%12\relax}%
\providecommand \@@startlink[1]{}%
\providecommand \@@endlink[0]{}%
\providecommand \url  [0]{\begingroup\@sanitize@url \@url }%
\providecommand \@url [1]{\endgroup\@href {#1}{\urlprefix }}%
\providecommand \urlprefix  [0]{URL }%
\providecommand \Eprint [0]{\href }%
\providecommand \doibase [0]{https://doi.org/}%
\providecommand \selectlanguage [0]{\@gobble}%
\providecommand \bibinfo  [0]{\@secondoftwo}%
\providecommand \bibfield  [0]{\@secondoftwo}%
\providecommand \translation [1]{[#1]}%
\providecommand \BibitemOpen [0]{}%
\providecommand \bibitemStop [0]{}%
\providecommand \bibitemNoStop [0]{.\EOS\space}%
\providecommand \EOS [0]{\spacefactor3000\relax}%
\providecommand \BibitemShut  [1]{\csname bibitem#1\endcsname}%
\let\auto@bib@innerbib\@empty
\bibitem [{\citenamefont {Breunig}(2016)}]{Breunig16}%
  \BibitemOpen
  \bibfield  {author} {\bibinfo {author} {\bibfnamefont {I.}~\bibnamefont
  {Breunig}},\ }\bibfield  {title} {\bibinfo {title} {Three-wave mixing in
  whispering gallery resonators},\ }\href
  {https://doi.org/10.1002/lpor.201600038} {\bibfield  {journal} {\bibinfo
  {journal} {Laser Photonics Rev.}\ }\textbf {\bibinfo {volume} {10}},\
  \bibinfo {pages} {569} (\bibinfo {year} {2016})}\BibitemShut {NoStop}%
\bibitem [{\citenamefont {Strekalov}\ \emph {et~al.}(2016)\citenamefont
  {Strekalov}, \citenamefont {Marquardt}, \citenamefont {Matsko}, \citenamefont
  {Schwefel},\ and\ \citenamefont {Leuchs}}]{Strekalov16}%
  \BibitemOpen
  \bibfield  {author} {\bibinfo {author} {\bibfnamefont {D.~V.}\ \bibnamefont
  {Strekalov}}, \bibinfo {author} {\bibfnamefont {C.}~\bibnamefont
  {Marquardt}}, \bibinfo {author} {\bibfnamefont {A.~B.}\ \bibnamefont
  {Matsko}}, \bibinfo {author} {\bibfnamefont {H.~G.~L.}\ \bibnamefont
  {Schwefel}},\ and\ \bibinfo {author} {\bibfnamefont {G.}~\bibnamefont
  {Leuchs}},\ }\bibfield  {title} {\bibinfo {title} {Nonlinear and quantum
  optics with whispering gallery resonators},\ }\href
  {https://doi.org/10.1088/2040-8978/18/12/123002} {\bibfield  {journal}
  {\bibinfo  {journal} {J. Opt.}\ }\textbf {\bibinfo {volume} {18}},\ \bibinfo
  {pages} {123002} (\bibinfo {year} {2016})}\BibitemShut {NoStop}%
\bibitem [{\citenamefont {Ilchenko}\ \emph {et~al.}(2004)\citenamefont
  {Ilchenko}, \citenamefont {Savchenkov}, \citenamefont {Matsko},\ and\
  \citenamefont {Maleki}}]{Ilchenko04}%
  \BibitemOpen
  \bibfield  {author} {\bibinfo {author} {\bibfnamefont {V.~S.}\ \bibnamefont
  {Ilchenko}}, \bibinfo {author} {\bibfnamefont {A.~A.}\ \bibnamefont
  {Savchenkov}}, \bibinfo {author} {\bibfnamefont {A.~B.}\ \bibnamefont
  {Matsko}},\ and\ \bibinfo {author} {\bibfnamefont {L.}~\bibnamefont
  {Maleki}},\ }\bibfield  {title} {\bibinfo {title} {Nonlinear optics and
  crystalline whispering gallery mode cavities},\ }\href
  {https://doi.org/10.1103/PhysRevLett.92.043903} {\bibfield  {journal}
  {\bibinfo  {journal} {Phys. Rev. Lett.}\ }\textbf {\bibinfo {volume} {92}},\
  \bibinfo {pages} {043903} (\bibinfo {year} {2004})}\BibitemShut {NoStop}%
\bibitem [{\citenamefont {Szabados}\ \emph
  {et~al.}(2020{\natexlab{a}})\citenamefont {Szabados}, \citenamefont
  {Puzyrev}, \citenamefont {Minet}, \citenamefont {Reis}, \citenamefont {Buse},
  \citenamefont {Villois}, \citenamefont {Skryabin},\ and\ \citenamefont
  {Breunig}}]{Szabados20}%
  \BibitemOpen
  \bibfield  {author} {\bibinfo {author} {\bibfnamefont {J.}~\bibnamefont
  {Szabados}}, \bibinfo {author} {\bibfnamefont {D.~N.}\ \bibnamefont
  {Puzyrev}}, \bibinfo {author} {\bibfnamefont {Y.}~\bibnamefont {Minet}},
  \bibinfo {author} {\bibfnamefont {L.}~\bibnamefont {Reis}}, \bibinfo {author}
  {\bibfnamefont {K.}~\bibnamefont {Buse}}, \bibinfo {author} {\bibfnamefont
  {A.}~\bibnamefont {Villois}}, \bibinfo {author} {\bibfnamefont {D.~V.}\
  \bibnamefont {Skryabin}},\ and\ \bibinfo {author} {\bibfnamefont
  {I.}~\bibnamefont {Breunig}},\ }\bibfield  {title} {\bibinfo {title}
  {Frequency comb generation via cascaded second-order nonlinearities in
  microresonators},\ }\href {https://doi.org/10.1103/PhysRevLett.124.203902}
  {\bibfield  {journal} {\bibinfo  {journal} {Phys. Rev. Lett.}\ }\textbf
  {\bibinfo {volume} {124}},\ \bibinfo {pages} {203902} (\bibinfo {year}
  {2020}{\natexlab{a}})}\BibitemShut {NoStop}%
\bibitem [{\citenamefont {Hendry}\ \emph {et~al.}(2020)\citenamefont {Hendry},
  \citenamefont {Trainor}, \citenamefont {Xu}, \citenamefont {Coen},
  \citenamefont {Murdoch}, \citenamefont {Schwefel},\ and\ \citenamefont
  {Erkintalo}}]{Hendry20}%
  \BibitemOpen
  \bibfield  {author} {\bibinfo {author} {\bibfnamefont {I.}~\bibnamefont
  {Hendry}}, \bibinfo {author} {\bibfnamefont {L.~S.}\ \bibnamefont {Trainor}},
  \bibinfo {author} {\bibfnamefont {Y.}~\bibnamefont {Xu}}, \bibinfo {author}
  {\bibfnamefont {S.}~\bibnamefont {Coen}}, \bibinfo {author} {\bibfnamefont
  {S.~G.}\ \bibnamefont {Murdoch}}, \bibinfo {author} {\bibfnamefont
  {H.~G.~L.}\ \bibnamefont {Schwefel}},\ and\ \bibinfo {author} {\bibfnamefont
  {M.}~\bibnamefont {Erkintalo}},\ }\bibfield  {title} {\bibinfo {title}
  {Experimental observation of internally pumped parametric oscillation and
  quadratic comb generation in a $\chi^{(2)}$ whispering-gallery-mode
  microresonator},\ }\href {https://doi.org/10.1364/OL.385751} {\bibfield
  {journal} {\bibinfo  {journal} {Opt. Lett.}\ }\textbf {\bibinfo {volume}
  {45}},\ \bibinfo {pages} {1204} (\bibinfo {year} {2020})}\BibitemShut
  {NoStop}%
\bibitem [{\citenamefont {Szabados}\ \emph
  {et~al.}(2020{\natexlab{b}})\citenamefont {Szabados}, \citenamefont
  {Sturman},\ and\ \citenamefont {Breunig}}]{Szabados20APLP}%
  \BibitemOpen
  \bibfield  {author} {\bibinfo {author} {\bibfnamefont {J.}~\bibnamefont
  {Szabados}}, \bibinfo {author} {\bibfnamefont {B.}~\bibnamefont {Sturman}},\
  and\ \bibinfo {author} {\bibfnamefont {I.}~\bibnamefont {Breunig}},\
  }\bibfield  {title} {\bibinfo {title} {Frequency comb generation threshold
  via second-harmonic excitation in $\chi^{(2)}$ optical microresonators},\
  }\href {https://doi.org/10.1063/5.0021424} {\bibfield  {journal} {\bibinfo
  {journal} {APL Photonics}\ }\textbf {\bibinfo {volume} {5}},\ \bibinfo
  {pages} {116102} (\bibinfo {year} {2020}{\natexlab{b}})}\BibitemShut
  {NoStop}%
\bibitem [{\citenamefont {Sturman}\ and\ \citenamefont
  {Breunig}(2011)}]{Sturman11}%
  \BibitemOpen
  \bibfield  {author} {\bibinfo {author} {\bibfnamefont {B.}~\bibnamefont
  {Sturman}}\ and\ \bibinfo {author} {\bibfnamefont {I.}~\bibnamefont
  {Breunig}},\ }\bibfield  {title} {\bibinfo {title} {Generic description of
  second-order nonlinear phenomena in whispering-gallery resonators},\ }\href
  {https://doi.org/10.1364/JOSAB.28.002465} {\bibfield  {journal} {\bibinfo
  {journal} {J. Opt. Soc. Am. B}\ }\textbf {\bibinfo {volume} {28}},\ \bibinfo
  {pages} {2465} (\bibinfo {year} {2011})}\BibitemShut {NoStop}%
\bibitem [{\citenamefont {Sturman}\ \emph {et~al.}(2012)\citenamefont
  {Sturman}, \citenamefont {Beckmann},\ and\ \citenamefont
  {Breunig}}]{Sturman12}%
  \BibitemOpen
  \bibfield  {author} {\bibinfo {author} {\bibfnamefont {B.}~\bibnamefont
  {Sturman}}, \bibinfo {author} {\bibfnamefont {T.}~\bibnamefont {Beckmann}},\
  and\ \bibinfo {author} {\bibfnamefont {I.}~\bibnamefont {Breunig}},\
  }\bibfield  {title} {\bibinfo {title} {Quasi-resonant and quasi-phase-matched
  nonlinear second-order phenomena in whispering-gallery resonators},\ }\href
  {https://doi.org/10.1364/JOSAB.29.003087} {\bibfield  {journal} {\bibinfo
  {journal} {J. Opt. Soc. Am. B}\ }\textbf {\bibinfo {volume} {29}},\ \bibinfo
  {pages} {3087} (\bibinfo {year} {2012})}\BibitemShut {NoStop}%
\bibitem [{\citenamefont {Ricciardi}\ \emph {et~al.}(2015)\citenamefont
  {Ricciardi}, \citenamefont {Mosca}, \citenamefont {Parisi}, \citenamefont
  {Maddaloni}, \citenamefont {Santamaria}, \citenamefont {{De Natale}},\ and\
  \citenamefont {{De Rosa}}}]{Ricciardi15}%
  \BibitemOpen
  \bibfield  {author} {\bibinfo {author} {\bibfnamefont {I.}~\bibnamefont
  {Ricciardi}}, \bibinfo {author} {\bibfnamefont {S.}~\bibnamefont {Mosca}},
  \bibinfo {author} {\bibfnamefont {M.}~\bibnamefont {Parisi}}, \bibinfo
  {author} {\bibfnamefont {P.}~\bibnamefont {Maddaloni}}, \bibinfo {author}
  {\bibfnamefont {L.}~\bibnamefont {Santamaria}}, \bibinfo {author}
  {\bibfnamefont {P.}~\bibnamefont {{De Natale}}},\ and\ \bibinfo {author}
  {\bibfnamefont {M.}~\bibnamefont {{De Rosa}}},\ }\bibfield  {title} {\bibinfo
  {title} {Frequency comb generation in quadratic nonlinear media},\ }\href
  {https://doi.org/10.1103/PhysRevA.91.063839} {\bibfield  {journal} {\bibinfo
  {journal} {Phys. Rev. A}\ }\textbf {\bibinfo {volume} {91}},\ \bibinfo
  {pages} {063839} (\bibinfo {year} {2015})}\BibitemShut {NoStop}%
\bibitem [{\citenamefont {Leo}\ \emph {et~al.}(2016{\natexlab{a}})\citenamefont
  {Leo}, \citenamefont {Hansson}, \citenamefont {Ricciardi}, \citenamefont {{De
  Rosa}}, \citenamefont {Coen}, \citenamefont {Wabnitz},\ and\ \citenamefont
  {Erkintalo}}]{Leo16}%
  \BibitemOpen
  \bibfield  {author} {\bibinfo {author} {\bibfnamefont {F.}~\bibnamefont
  {Leo}}, \bibinfo {author} {\bibfnamefont {T.}~\bibnamefont {Hansson}},
  \bibinfo {author} {\bibfnamefont {I.}~\bibnamefont {Ricciardi}}, \bibinfo
  {author} {\bibfnamefont {M.}~\bibnamefont {{De Rosa}}}, \bibinfo {author}
  {\bibfnamefont {S.}~\bibnamefont {Coen}}, \bibinfo {author} {\bibfnamefont
  {S.}~\bibnamefont {Wabnitz}},\ and\ \bibinfo {author} {\bibfnamefont
  {M.}~\bibnamefont {Erkintalo}},\ }\bibfield  {title} {\bibinfo {title}
  {Walk-off-induced modulation instability, temporal pattern formation, and
  frequency comb generation in cavity-enhanced second-harmonic generation},\
  }\href {https://doi.org/10.1103/PhysRevLett.116.033901} {\bibfield  {journal}
  {\bibinfo  {journal} {Phys. Rev. Lett.}\ }\textbf {\bibinfo {volume} {116}},\
  \bibinfo {pages} {033901} (\bibinfo {year} {2016}{\natexlab{a}})}\BibitemShut
  {NoStop}%
\bibitem [{\citenamefont {Leo}\ \emph {et~al.}(2016{\natexlab{b}})\citenamefont
  {Leo}, \citenamefont {Hansson}, \citenamefont {Ricciardi}, \citenamefont {{De
  Rosa}}, \citenamefont {Coen}, \citenamefont {Wabnitz},\ and\ \citenamefont
  {Erkintalo}}]{Leo16pra}%
  \BibitemOpen
  \bibfield  {author} {\bibinfo {author} {\bibfnamefont {F.}~\bibnamefont
  {Leo}}, \bibinfo {author} {\bibfnamefont {T.}~\bibnamefont {Hansson}},
  \bibinfo {author} {\bibfnamefont {I.}~\bibnamefont {Ricciardi}}, \bibinfo
  {author} {\bibfnamefont {M.}~\bibnamefont {{De Rosa}}}, \bibinfo {author}
  {\bibfnamefont {S.}~\bibnamefont {Coen}}, \bibinfo {author} {\bibfnamefont
  {S.}~\bibnamefont {Wabnitz}},\ and\ \bibinfo {author} {\bibfnamefont
  {M.}~\bibnamefont {Erkintalo}},\ }\bibfield  {title} {\bibinfo {title}
  {Frequency-comb formation in doubly resonant second-harmonic generation},\
  }\href {https://doi.org/10.1103/PhysRevA.93.043831} {\bibfield  {journal}
  {\bibinfo  {journal} {Phys. Rev. A}\ }\textbf {\bibinfo {volume} {93}},\
  \bibinfo {pages} {043831} (\bibinfo {year} {2016}{\natexlab{b}})}\BibitemShut
  {NoStop}%
\bibitem [{\citenamefont {Mosca}\ \emph {et~al.}(2018)\citenamefont {Mosca},
  \citenamefont {Parisi}, \citenamefont {Ricciardi}, \citenamefont {Leo},
  \citenamefont {Hansson}, \citenamefont {Erkintalo}, \citenamefont
  {Maddaloni}, \citenamefont {{De Natale}}, \citenamefont {Wabnitz},\ and\
  \citenamefont {{De Rosa}}}]{Mosca18}%
  \BibitemOpen
  \bibfield  {author} {\bibinfo {author} {\bibfnamefont {S.}~\bibnamefont
  {Mosca}}, \bibinfo {author} {\bibfnamefont {M.}~\bibnamefont {Parisi}},
  \bibinfo {author} {\bibfnamefont {I.}~\bibnamefont {Ricciardi}}, \bibinfo
  {author} {\bibfnamefont {F.}~\bibnamefont {Leo}}, \bibinfo {author}
  {\bibfnamefont {T.}~\bibnamefont {Hansson}}, \bibinfo {author} {\bibfnamefont
  {M.}~\bibnamefont {Erkintalo}}, \bibinfo {author} {\bibfnamefont
  {P.}~\bibnamefont {Maddaloni}}, \bibinfo {author} {\bibfnamefont
  {P.}~\bibnamefont {{De Natale}}}, \bibinfo {author} {\bibfnamefont
  {S.}~\bibnamefont {Wabnitz}},\ and\ \bibinfo {author} {\bibfnamefont
  {M.}~\bibnamefont {{De Rosa}}},\ }\bibfield  {title} {\bibinfo {title}
  {Modulation instability induced frequency comb generation in a continuously
  pumped optical parametric oscillator},\ }\href
  {https://doi.org/10.1103/PhysRevLett.121.093903} {\bibfield  {journal}
  {\bibinfo  {journal} {Phys. Rev. Lett.}\ }\textbf {\bibinfo {volume} {121}},\
  \bibinfo {pages} {093903} (\bibinfo {year} {2018})}\BibitemShut {NoStop}%
\bibitem [{\citenamefont {Villois}\ and\ \citenamefont
  {Skryabin}(2019)}]{Villois19b}%
  \BibitemOpen
  \bibfield  {author} {\bibinfo {author} {\bibfnamefont {A.}~\bibnamefont
  {Villois}}\ and\ \bibinfo {author} {\bibfnamefont {D.~V.}\ \bibnamefont
  {Skryabin}},\ }\bibfield  {title} {\bibinfo {title} {Soliton and
  quasi-soliton frequency combs due to second harmonic generation in
  microresonators},\ }\href {https://doi.org/10.1364/OE.27.007098} {\bibfield
  {journal} {\bibinfo  {journal} {Opt. Express}\ }\textbf {\bibinfo {volume}
  {27}},\ \bibinfo {pages} {7098} (\bibinfo {year} {2019})}\BibitemShut
  {NoStop}%
\bibitem [{\citenamefont {Smirnov}\ \emph {et~al.}(2020)\citenamefont
  {Smirnov}, \citenamefont {Sturman}, \citenamefont {Podivilov},\ and\
  \citenamefont {Breunig}}]{Smirnov20}%
  \BibitemOpen
  \bibfield  {author} {\bibinfo {author} {\bibfnamefont {S.}~\bibnamefont
  {Smirnov}}, \bibinfo {author} {\bibfnamefont {B.}~\bibnamefont {Sturman}},
  \bibinfo {author} {\bibfnamefont {E.}~\bibnamefont {Podivilov}},\ and\
  \bibinfo {author} {\bibfnamefont {I.}~\bibnamefont {Breunig}},\ }\bibfield
  {title} {\bibinfo {title} {Walk-off controlled self-starting frequency combs
  in $\chi^{(2)}$ optical microresonators},\ }\href
  {https://doi.org/10.1364/OE.395360} {\bibfield  {journal} {\bibinfo
  {journal} {Opt. Express}\ }\textbf {\bibinfo {volume} {28}},\ \bibinfo
  {pages} {18006} (\bibinfo {year} {2020})}\BibitemShut {NoStop}%
\bibitem [{\citenamefont {Lobanov}\ \emph {et~al.}(2020)\citenamefont
  {Lobanov}, \citenamefont {Kondratiev}, \citenamefont {Shitikov},\ and\
  \citenamefont {Bilenko}}]{Lobanov20}%
  \BibitemOpen
  \bibfield  {author} {\bibinfo {author} {\bibfnamefont {V.~E.}\ \bibnamefont
  {Lobanov}}, \bibinfo {author} {\bibfnamefont {N.~M.}\ \bibnamefont
  {Kondratiev}}, \bibinfo {author} {\bibfnamefont {A.~E.}\ \bibnamefont
  {Shitikov}},\ and\ \bibinfo {author} {\bibfnamefont {I.~A.}\ \bibnamefont
  {Bilenko}},\ }\bibfield  {title} {\bibinfo {title} {Two-color flat-top
  solitonic pulses in $\chi^{(2)}$ optical microresonators via second-harmonic
  generation},\ }\href {https://doi.org/10.1103/PhysRevA.101.013831} {\bibfield
   {journal} {\bibinfo  {journal} {Phys. Rev. A}\ }\textbf {\bibinfo {volume}
  {101}},\ \bibinfo {pages} {013831} (\bibinfo {year} {2020})}\BibitemShut
  {NoStop}%
\bibitem [{\citenamefont {Breunig}\ \emph {et~al.}(2013)\citenamefont
  {Breunig}, \citenamefont {Sturman}, \citenamefont {B\"uckle}, \citenamefont
  {Werner},\ and\ \citenamefont {Buse}}]{Breunig13}%
  \BibitemOpen
  \bibfield  {author} {\bibinfo {author} {\bibfnamefont {I.}~\bibnamefont
  {Breunig}}, \bibinfo {author} {\bibfnamefont {B.}~\bibnamefont {Sturman}},
  \bibinfo {author} {\bibfnamefont {A.}~\bibnamefont {B\"uckle}}, \bibinfo
  {author} {\bibfnamefont {C.~S.}\ \bibnamefont {Werner}},\ and\ \bibinfo
  {author} {\bibfnamefont {K.}~\bibnamefont {Buse}},\ }\bibfield  {title}
  {\bibinfo {title} {Structure of pump resonances during optical parametric
  oscillation in whispering gallery resonators},\ }\href
  {https://doi.org/10.1364/OL.38.003316} {\bibfield  {journal} {\bibinfo
  {journal} {Opt. Lett.}\ }\textbf {\bibinfo {volume} {38}},\ \bibinfo {pages}
  {3316} (\bibinfo {year} {2013})}\BibitemShut {NoStop}%
\bibitem [{\citenamefont {F\"urst}\ \emph {et~al.}(2010)\citenamefont
  {F\"urst}, \citenamefont {Strekalov}, \citenamefont {Elser}, \citenamefont
  {Lassen}, \citenamefont {Andersen}, \citenamefont {Marquardt},\ and\
  \citenamefont {Leuchs}}]{Fuerst10shg}%
  \BibitemOpen
  \bibfield  {author} {\bibinfo {author} {\bibfnamefont {J.~U.}\ \bibnamefont
  {F\"urst}}, \bibinfo {author} {\bibfnamefont {D.~V.}\ \bibnamefont
  {Strekalov}}, \bibinfo {author} {\bibfnamefont {D.}~\bibnamefont {Elser}},
  \bibinfo {author} {\bibfnamefont {M.}~\bibnamefont {Lassen}}, \bibinfo
  {author} {\bibfnamefont {U.~L.}\ \bibnamefont {Andersen}}, \bibinfo {author}
  {\bibfnamefont {C.}~\bibnamefont {Marquardt}},\ and\ \bibinfo {author}
  {\bibfnamefont {G.}~\bibnamefont {Leuchs}},\ }\bibfield  {title} {\bibinfo
  {title} {Naturally phase-matched second-harmonic generation in a
  whispering-gallery-mode resonator},\ }\href
  {https://doi.org/10.1103/PhysRevLett.104.153901} {\bibfield  {journal}
  {\bibinfo  {journal} {Phys. Rev. Lett.}\ }\textbf {\bibinfo {volume} {104}},\
  \bibinfo {pages} {153901} (\bibinfo {year} {2010})}\BibitemShut {NoStop}%
\bibitem [{\citenamefont {Yariv}(2000)}]{Yariv2000}%
  \BibitemOpen
  \bibfield  {author} {\bibinfo {author} {\bibfnamefont {A.}~\bibnamefont
  {Yariv}},\ }\bibfield  {title} {\bibinfo {title} {Universal relations for
  coupling of optical power between microresonators and dielectric
  waveguides},\ }\href {https://doi.org/10.1049/el:20000340} {\bibfield
  {journal} {\bibinfo  {journal} {Electron. Lett.}\ }\textbf {\bibinfo {volume}
  {36}},\ \bibinfo {pages} {321} (\bibinfo {year} {2000})}\BibitemShut
  {NoStop}%
\bibitem [{\citenamefont {Yariv}(2002)}]{Yariv2002}%
  \BibitemOpen
  \bibfield  {author} {\bibinfo {author} {\bibfnamefont {A.}~\bibnamefont
  {Yariv}},\ }\bibfield  {title} {\bibinfo {title} {Critical coupling and its
  control in optical waveguide-ring resonator systems},\ }\href
  {https://doi.org/10.1109/68.992585} {\bibfield  {journal} {\bibinfo
  {journal} {IEEE Photon. Technol. Lett.}\ }\textbf {\bibinfo {volume} {14}},\
  \bibinfo {pages} {483} (\bibinfo {year} {2002})}\BibitemShut {NoStop}%
\bibitem [{\citenamefont {F\"urst}\ \emph {et~al.}(2015)\citenamefont
  {F\"urst}, \citenamefont {Buse}, \citenamefont {Breunig}, \citenamefont
  {Becker}, \citenamefont {Liebertz},\ and\ \citenamefont
  {Bohat\'y}}]{Fuerst15}%
  \BibitemOpen
  \bibfield  {author} {\bibinfo {author} {\bibfnamefont {J.~U.}\ \bibnamefont
  {F\"urst}}, \bibinfo {author} {\bibfnamefont {K.}~\bibnamefont {Buse}},
  \bibinfo {author} {\bibfnamefont {I.}~\bibnamefont {Breunig}}, \bibinfo
  {author} {\bibfnamefont {P.}~\bibnamefont {Becker}}, \bibinfo {author}
  {\bibfnamefont {J.}~\bibnamefont {Liebertz}},\ and\ \bibinfo {author}
  {\bibfnamefont {L.}~\bibnamefont {Bohat\'y}},\ }\bibfield  {title} {\bibinfo
  {title} {Second-harmonic generation of light at 245 nm in a lithium
  tetraborate whispering gallery resonator},\ }\href
  {https://doi.org/10.1364/OL.40.001932} {\bibfield  {journal} {\bibinfo
  {journal} {Opt. Lett.}\ }\textbf {\bibinfo {volume} {40}},\ \bibinfo {pages}
  {1932} (\bibinfo {year} {2015})}\BibitemShut {NoStop}%
\bibitem [{\citenamefont {Minet}\ \emph {et~al.}(2020)\citenamefont {Minet},
  \citenamefont {Reis}, \citenamefont {Szabados}, \citenamefont {Werner},
  \citenamefont {Zappe}, \citenamefont {Buse},\ and\ \citenamefont
  {Breunig}}]{Minet2020}%
  \BibitemOpen
  \bibfield  {author} {\bibinfo {author} {\bibfnamefont {Y.}~\bibnamefont
  {Minet}}, \bibinfo {author} {\bibfnamefont {L.}~\bibnamefont {Reis}},
  \bibinfo {author} {\bibfnamefont {J.}~\bibnamefont {Szabados}}, \bibinfo
  {author} {\bibfnamefont {C.~S.}\ \bibnamefont {Werner}}, \bibinfo {author}
  {\bibfnamefont {H.}~\bibnamefont {Zappe}}, \bibinfo {author} {\bibfnamefont
  {K.}~\bibnamefont {Buse}},\ and\ \bibinfo {author} {\bibfnamefont
  {I.}~\bibnamefont {Breunig}},\ }\bibfield  {title} {\bibinfo {title}
  {Pockels-effect-based adiabatic frequency conversion in ultrahigh-{$Q$}
  microresonators},\ }\href {https://doi.org/10.1364/OE.378112} {\bibfield
  {journal} {\bibinfo  {journal} {Opt. Express}\ }\textbf {\bibinfo {volume}
  {28}},\ \bibinfo {pages} {2939} (\bibinfo {year} {2020})}\BibitemShut
  {NoStop}%
\bibitem [{\citenamefont {Wei}\ \emph {et~al.}(2018)\citenamefont {Wei},
  \citenamefont {Murray}, \citenamefont {Hopkins}, \citenamefont {Krein},
  \citenamefont {Zawilski}, \citenamefont {Schunemann},\ and\ \citenamefont
  {Guha}}]{Wei18}%
  \BibitemOpen
  \bibfield  {author} {\bibinfo {author} {\bibfnamefont {J.}~\bibnamefont
  {Wei}}, \bibinfo {author} {\bibfnamefont {J.~M.}\ \bibnamefont {Murray}},
  \bibinfo {author} {\bibfnamefont {F.~K.}\ \bibnamefont {Hopkins}}, \bibinfo
  {author} {\bibfnamefont {D.~M.}\ \bibnamefont {Krein}}, \bibinfo {author}
  {\bibfnamefont {K.~T.}\ \bibnamefont {Zawilski}}, \bibinfo {author}
  {\bibfnamefont {P.~G.}\ \bibnamefont {Schunemann}},\ and\ \bibinfo {author}
  {\bibfnamefont {S.}~\bibnamefont {Guha}},\ }\bibfield  {title} {\bibinfo
  {title} {Measurement of refractive indices of {CdSiP}$_{2}$ at temperatures
  from 90 to 450 {K}},\ }\href {https://doi.org/10.1364/OME.8.000235}
  {\bibfield  {journal} {\bibinfo  {journal} {Opt. Mater. Express}\ }\textbf
  {\bibinfo {volume} {8}},\ \bibinfo {pages} {235} (\bibinfo {year}
  {2018})}\BibitemShut {NoStop}%
\bibitem [{\citenamefont {Gorodetsky}\ and\ \citenamefont
  {Fomin}(2006)}]{Gorodetsky06}%
  \BibitemOpen
  \bibfield  {author} {\bibinfo {author} {\bibfnamefont {M.~L.}\ \bibnamefont
  {Gorodetsky}}\ and\ \bibinfo {author} {\bibfnamefont {A.~E.}\ \bibnamefont
  {Fomin}},\ }\bibfield  {title} {\bibinfo {title} {Geometrical theory of
  whispering-gallery modes},\ }\href
  {https://doi.org/10.1109/JSTQE.2005.862954} {\bibfield  {journal} {\bibinfo
  {journal} {IEEE J. Sel. Top. Quantum Electron.}\ }\textbf {\bibinfo {volume}
  {12}},\ \bibinfo {pages} {33} (\bibinfo {year} {2006})}\BibitemShut {NoStop}%
\bibitem [{\citenamefont {Jia}\ \emph {et~al.}(2018)\citenamefont {Jia},
  \citenamefont {Hanka}, \citenamefont {Zawilski}, \citenamefont {Schunemann},
  \citenamefont {Buse},\ and\ \citenamefont {Breunig}}]{Jia18}%
  \BibitemOpen
  \bibfield  {author} {\bibinfo {author} {\bibfnamefont {Y.}~\bibnamefont
  {Jia}}, \bibinfo {author} {\bibfnamefont {K.}~\bibnamefont {Hanka}}, \bibinfo
  {author} {\bibfnamefont {K.~T.}\ \bibnamefont {Zawilski}}, \bibinfo {author}
  {\bibfnamefont {P.~G.}\ \bibnamefont {Schunemann}}, \bibinfo {author}
  {\bibfnamefont {K.}~\bibnamefont {Buse}},\ and\ \bibinfo {author}
  {\bibfnamefont {I.}~\bibnamefont {Breunig}},\ }\bibfield  {title} {\bibinfo
  {title} {Continuous-wave whispering-gallery optical parametric oscillator
  based on {CdSiP}$_{2}$},\ }\href {https://doi.org/10.1364/OE.26.010833}
  {\bibfield  {journal} {\bibinfo  {journal} {Opt. Express}\ }\textbf {\bibinfo
  {volume} {26}},\ \bibinfo {pages} {10833} (\bibinfo {year}
  {2018})}\BibitemShut {NoStop}%
\bibitem [{\citenamefont {Leidinger}\ \emph {et~al.}(2015)\citenamefont
  {Leidinger}, \citenamefont {Fieberg}, \citenamefont {Waasem}, \citenamefont
  {K\"uhnemann}, \citenamefont {Buse},\ and\ \citenamefont
  {Breunig}}]{Leidinger15}%
  \BibitemOpen
  \bibfield  {author} {\bibinfo {author} {\bibfnamefont {M.}~\bibnamefont
  {Leidinger}}, \bibinfo {author} {\bibfnamefont {S.}~\bibnamefont {Fieberg}},
  \bibinfo {author} {\bibfnamefont {N.}~\bibnamefont {Waasem}}, \bibinfo
  {author} {\bibfnamefont {F.}~\bibnamefont {K\"uhnemann}}, \bibinfo {author}
  {\bibfnamefont {K.}~\bibnamefont {Buse}},\ and\ \bibinfo {author}
  {\bibfnamefont {I.}~\bibnamefont {Breunig}},\ }\bibfield  {title} {\bibinfo
  {title} {Comparative study on three highly sensitive absorption measurement
  techniques characterizing lithium niobate over its entire transparent
  spectral range},\ }\href {https://doi.org/10.1364/OE.23.021690} {\bibfield
  {journal} {\bibinfo  {journal} {Opt. Express}\ }\textbf {\bibinfo {volume}
  {23}},\ \bibinfo {pages} {21690} (\bibinfo {year} {2015})}\BibitemShut
  {NoStop}%
\bibitem [{\citenamefont {Umemura}\ and\ \citenamefont
  {Matsuda}(2016)}]{Umemura16}%
  \BibitemOpen
  \bibfield  {author} {\bibinfo {author} {\bibfnamefont {N.}~\bibnamefont
  {Umemura}}\ and\ \bibinfo {author} {\bibfnamefont {D.}~\bibnamefont
  {Matsuda}},\ }\bibfield  {title} {\bibinfo {title} {Thermo-optic dispersion
  formula for the ordinary wave in 5 mol$\%$ {MgO} doped {LiNbO$_{3}$} and its
  application to temperature insensitive second-harmonic generation},\ }\href
  {https://doi.org/10.1016/j.optcom.2016.01.007} {\bibfield  {journal}
  {\bibinfo  {journal} {Opt. Commun.}\ }\textbf {\bibinfo {volume} {367}},\
  \bibinfo {pages} {167} (\bibinfo {year} {2016})}\BibitemShut {NoStop}%
\bibitem [{\citenamefont {M\'{e}ndez}\ \emph {et~al.}(1999)\citenamefont
  {M\'{e}ndez}, \citenamefont {{Garc\'{i}a-Caba\~{n}es}}, \citenamefont
  {Di\'{e}guez},\ and\ \citenamefont {Cabrera}}]{Mendez99}%
  \BibitemOpen
  \bibfield  {author} {\bibinfo {author} {\bibfnamefont {A.}~\bibnamefont
  {M\'{e}ndez}}, \bibinfo {author} {\bibfnamefont {A.}~\bibnamefont
  {{Garc\'{i}a-Caba\~{n}es}}}, \bibinfo {author} {\bibfnamefont
  {E.}~\bibnamefont {Di\'{e}guez}},\ and\ \bibinfo {author} {\bibfnamefont
  {J.~M.}\ \bibnamefont {Cabrera}},\ }\bibfield  {title} {\bibinfo {title}
  {Wavelength dependence of electro-optic coefficients in congruent and
  quasi-stoichiometric {LiNbO}$_{3}$},\ }\href
  {https://doi.org/10.1049/el:19990345} {\bibfield  {journal} {\bibinfo
  {journal} {Electron. Lett.}\ }\textbf {\bibinfo {volume} {35}},\ \bibinfo
  {pages} {498} (\bibinfo {year} {1999})}\BibitemShut {NoStop}%
\bibitem [{\citenamefont {Ilchenko}\ and\ \citenamefont
  {Gorodetskii}(1992)}]{Ilchenko92}%
  \BibitemOpen
  \bibfield  {author} {\bibinfo {author} {\bibfnamefont {V.~S.}\ \bibnamefont
  {Ilchenko}}\ and\ \bibinfo {author} {\bibfnamefont {M.~L.}\ \bibnamefont
  {Gorodetskii}},\ }\bibfield  {title} {\bibinfo {title} {Thermal nonlinear
  effects in optical whispering gallery microresonators},\ }\href@noop {}
  {\bibfield  {journal} {\bibinfo  {journal} {Laser Phys.}\ }\textbf {\bibinfo
  {volume} {2}},\ \bibinfo {pages} {1004} (\bibinfo {year} {1992})}\BibitemShut
  {NoStop}%
\bibitem [{\citenamefont {Carmon}\ \emph {et~al.}(2004)\citenamefont {Carmon},
  \citenamefont {Yang},\ and\ \citenamefont {Vahala}}]{Carmon04}%
  \BibitemOpen
  \bibfield  {author} {\bibinfo {author} {\bibfnamefont {T.}~\bibnamefont
  {Carmon}}, \bibinfo {author} {\bibfnamefont {L.}~\bibnamefont {Yang}},\ and\
  \bibinfo {author} {\bibfnamefont {K.~J.}\ \bibnamefont {Vahala}},\ }\bibfield
   {title} {\bibinfo {title} {Dynamical thermal behavior and thermal
  self-stability of microcavities},\ }\href
  {https://doi.org/10.1364/OPEX.12.004742} {\bibfield  {journal} {\bibinfo
  {journal} {Opt. Express}\ }\textbf {\bibinfo {volume} {12}},\ \bibinfo
  {pages} {4742} (\bibinfo {year} {2004})}\BibitemShut {NoStop}%
\bibitem [{\citenamefont {Ricciardi}\ \emph {et~al.}(2020)\citenamefont
  {Ricciardi}, \citenamefont {Mosca}, \citenamefont {Parisi}, \citenamefont
  {Leo}, \citenamefont {Hansson}, \citenamefont {Erkintalo}, \citenamefont
  {Maddaloni}, \citenamefont {{De Natale}}, \citenamefont {Wabnitz},\ and\
  \citenamefont {{De Rosa}}}]{Ricciardi20}%
  \BibitemOpen
  \bibfield  {author} {\bibinfo {author} {\bibfnamefont {I.}~\bibnamefont
  {Ricciardi}}, \bibinfo {author} {\bibfnamefont {S.}~\bibnamefont {Mosca}},
  \bibinfo {author} {\bibfnamefont {M.}~\bibnamefont {Parisi}}, \bibinfo
  {author} {\bibfnamefont {F.}~\bibnamefont {Leo}}, \bibinfo {author}
  {\bibfnamefont {T.}~\bibnamefont {Hansson}}, \bibinfo {author} {\bibfnamefont
  {M.}~\bibnamefont {Erkintalo}}, \bibinfo {author} {\bibfnamefont
  {P.}~\bibnamefont {Maddaloni}}, \bibinfo {author} {\bibfnamefont
  {P.}~\bibnamefont {{De Natale}}}, \bibinfo {author} {\bibfnamefont
  {S.}~\bibnamefont {Wabnitz}},\ and\ \bibinfo {author} {\bibfnamefont
  {M.}~\bibnamefont {{De Rosa}}},\ }\bibfield  {title} {\bibinfo {title}
  {Optical frequency combs in quadratically nonlinear resonators},\ }\href
  {https://doi.org/10.3390/mi11020230} {\bibfield  {journal} {\bibinfo
  {journal} {Micromachines}\ }\textbf {\bibinfo {volume} {11}},\ \bibinfo
  {pages} {230} (\bibinfo {year} {2020})}\BibitemShut {NoStop}%
\end{thebibliography}%


\providecommand{\noopsort}[1]{}\providecommand{\singleletter}[1]{#1}%
%






\end{document}